\documentclass[a4paper,11pt]{article}
\usepackage{jinstpub} 
\usepackage{lineno}
\usepackage{subcaption}
\usepackage{multirow}
\usepackage{hyperref}

\usepackage[printonlyused]{acronym}
\acrodef{pmt}[PMT]{photomultiplier tube}
\acrodef{hv}[HV]{high voltage}
\acrodef{shv}[SHV]{safe high voltage}
\acrodef{ac}[AC]{alternating current}
\acrodef{dc}[DC]{direct current}
\acrodef{pe}[PE]{photoelectron}
\acrodef{spe}[SPE]{single-photoelectron}
\acrodef{bugs}[BUGS]{Boulby UnderGround Screening}

\title{Characterisation of the Temperature-dependent Dark Rate of Hamamatsu R7081-100 10" Photomultiplier Tubes}

\author[a,1]{S. T. Wilson, \note{Corresponding author.}}
\author[a]{S. Fargher,}
\author[a]{R. Foster,}
\author[a]{M. Malek,}
\author[b]{M. Needham,}
\author[a]{A. Scarff,}
\author[b]{G. D. Smith,}
\affiliation[a]{Department of Physics and Astronomy, University of Sheffield, Hounsfield Road, S3 7RH, UK}
\affiliation[b]{School of Physics and Astronomy, University of Edinburgh, Mayfield Road, EH9 3JZ, UK}

\emailAdd{stephen.wilson@sheffield.ac.uk}

\abstract{Dark noise is a dominant background in \acp{pmt}, which are commonly used in liquid-filled particle detectors for single-photon detection to see the results of particle interactions. A major contribution to dark noise is thermionic emission from the photocathode.

The dark noise of Hamamatsu R7081-100 \acp{pmt} is characterised in a temperature and purity controlled water tank, with the thermionic emission contribution isolated. The results suggest that the intrinsic dark rate of \acp{pmt} does not depend on the medium, but does follow Richardson's law of thermionic emission. There are external contributions to the overall observed \ac{pmt} count rate identified, but the intrinsic \ac{pmt} dark rate in water matches that measured in air.}

\keywords{Photon detectors for UV, visible and IR photons (vacuum) (photomultipliers, HPDs, others), Cherenkov detectors, Neutrino detectors}

\begin{document}
\maketitle
\flushbottom

\section{Introduction}
\label{sec:intro}





\Acfp{pmt} are commonly used in particle detectors to detect photons resulting from particle interactions. In many detectors, particularly neutrino and dark matter detectors, the \acp{pmt} are submerged in a liquid medium such as pure water or liquid scintillator. However, they are typically characterised in air, normally at a single ambient temperature, before installation into a detector \cite{Zhang_2022,KAPTANOGLU201869,HEROLD2011S151,Brack_2013,AGUILAR2005132,CBauer_2011,ABBASI2010139}.

\Acp{pmt} used for single-photon detection have characteristic \ac{spe} pulses produced when a photon liberates an electron from the photocathode. However, photons are not the only mechanism by which electrons are liberated from the photocathode. Photons and other mechanisms are indistinguishable from the pulse, as they all cause electron emission making the pulses identical. A single electron will produce an \ac{spe} level pulse (a pulse with the same characteristics as a single-photon produced pulse).


Pulses produced in the absence of photons are referred to as ``dark" pulses, \acp{pmt} each have a characteristic count rate of these, and are a dominant background to single-photon detection. Thermionic emission, where an electron is liberated from the photocathode by thermal energy, is one of the major mechanisms that produces dark counts and should be characterised in detail.

It has been seen in detectors that the rate of pulses changes depending on the medium the \acp{pmt} are installed in and its conditions, e.g. in the LUX-ZEPLIN dark matter detector \cite{Wright2023}. As such, it is important to characterise PMTs in a realistic dielectric and magnetic environment, and with varying temperature to understand the performance and thermionic dark noise contribution that can be expected in a detector. Here, the characterisation of the temperature-dependent dark noise of 10" Hamamatsu R7081-100 PMTs when submerged in pure water is presented.
\section{Experimental setup and measurements}
\label{sec:setup_measurements}

\subsection{Photomultiplier tubes}
\label{subsec:pmts}

The \acp{pmt} used in these tests are low radioactivity Hamamatsu R7081-100 10" diameter tubes with a waterproof potting applied. An 80 m BELDEN YR53485 cable, connected by the vendor, is used to supply positive \ac{hv} and provide a return path for the signal as they are \ac{ac} coupled. The electronics in the \ac{pmt} base has an impedance of 50 $\Omega$.

\subsection{Electronics}
\label{subsec:electronics}

The \acp{pmt} are powered via a CAEN V6534P positive polarity \ac{hv} board. As the \acp{pmt} are \ac{ac} coupled, the \ac{ac} signal and \ac{dc} \ac{hv} need to be separated. A custom ``splitter" board, using a capacitor in a resistor-capacitor circuit as a high-pass filter, is used \cite{PMTQA} to block the \ac{dc} components and eliminate the \ac{dc} offset from the \ac{ac} signals. The splitter is designed to have the same impedance, 50 $\Omega$, as the electronics in the \ac{pmt} base to reduce signal reflections caused by impedance mismatches.


As the \ac{spe} signal from these \acp{pmt} has a mean amplitude of approximately 4 mV, the signal is amplified by a CAEN N979 10x fast amplifier. It is then digitised by a CAEN V1730B 500 MS/s digitizer where a 10 kHz external trigger is used to open a 220 ns acquisition window. The digitiser is read out via an optical link to a CAEN A2818 PCI CONET controller.

\subsection{Water tank}
\label{subsec:tank}

To characterise the \acp{pmt} in water, they need to be completely submerged. To achieve this, a 2000 L (2 m x 1 m x 1 m) 316L stainless steel water tank has been constructed at the University of Sheffield. The water in the tank is purified by a Veolia system to 18.2 M$\Omega\cdot$cm using a series of filters, a deionising cylinder and an ultraviolet lamp, and the temperature can be controlled in the 4 $^{\circ}$C to 30 $^{\circ}$C range by a heat exchange unit. The temperature of the water in the tank and purification system is measured by a series of temperature probes, and the resistivity of the water is measured as it enters and exits the purification system. The system is insulated, and the temperature is stable within 0.1 $^{\circ}$C in all but extreme ambient conditions. The resistivity of the water remains consistent at approximately 12.5 M$\Omega\cdot$cm in the tank during data taking periods.

\begin{figure}[!htb]
 \begin{subfigure}[b]{0.475\textwidth}
 \includegraphics[angle=270,width=\textwidth]{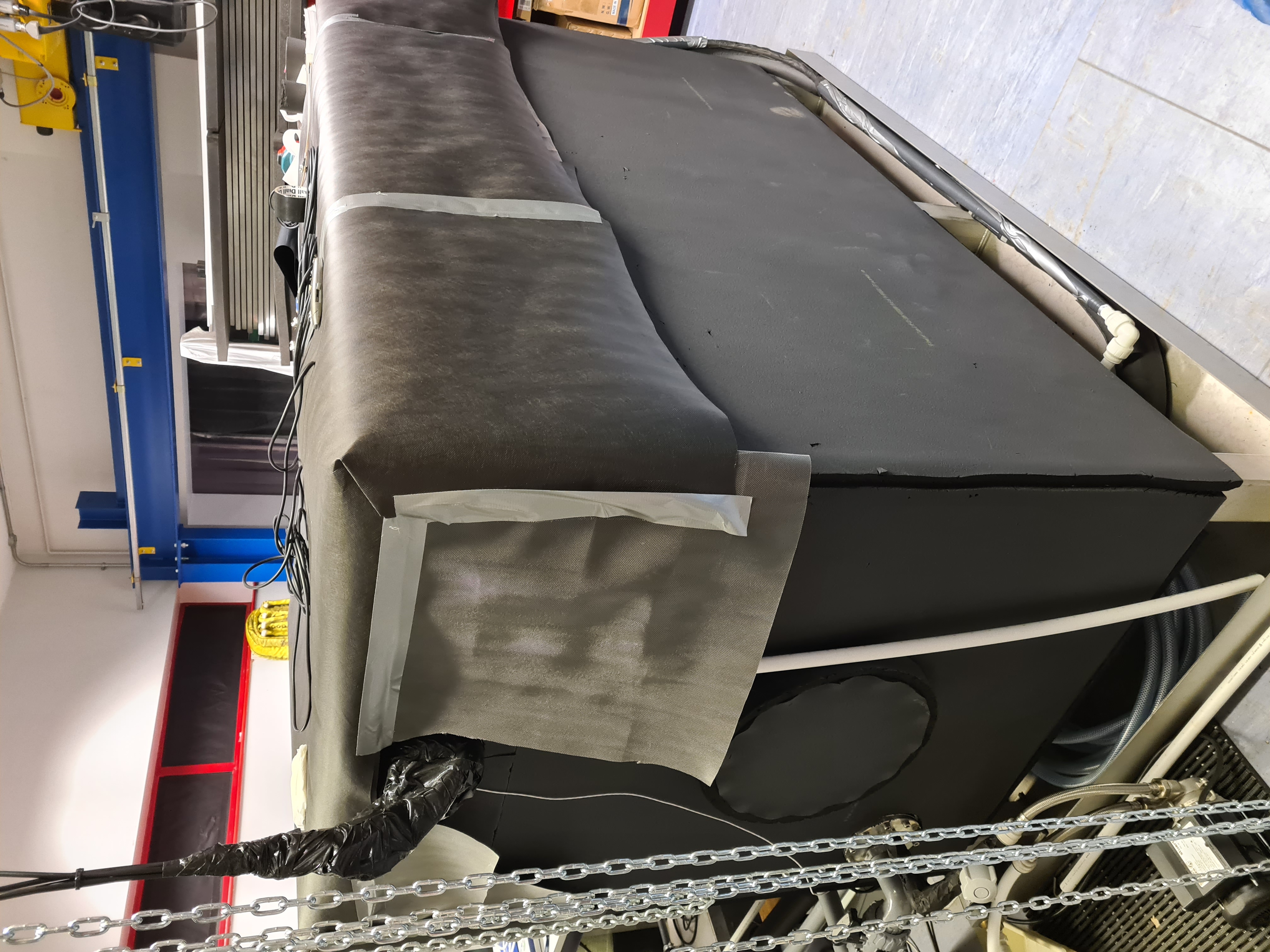}
 \caption{The 2000 L tank used to submerge \acp{pmt} in water.}
 \label{fig:tank}
 \end{subfigure}
 \hfill
 \begin{subfigure}[b]{0.475\textwidth}
 \includegraphics[angle=0,width=\textwidth]{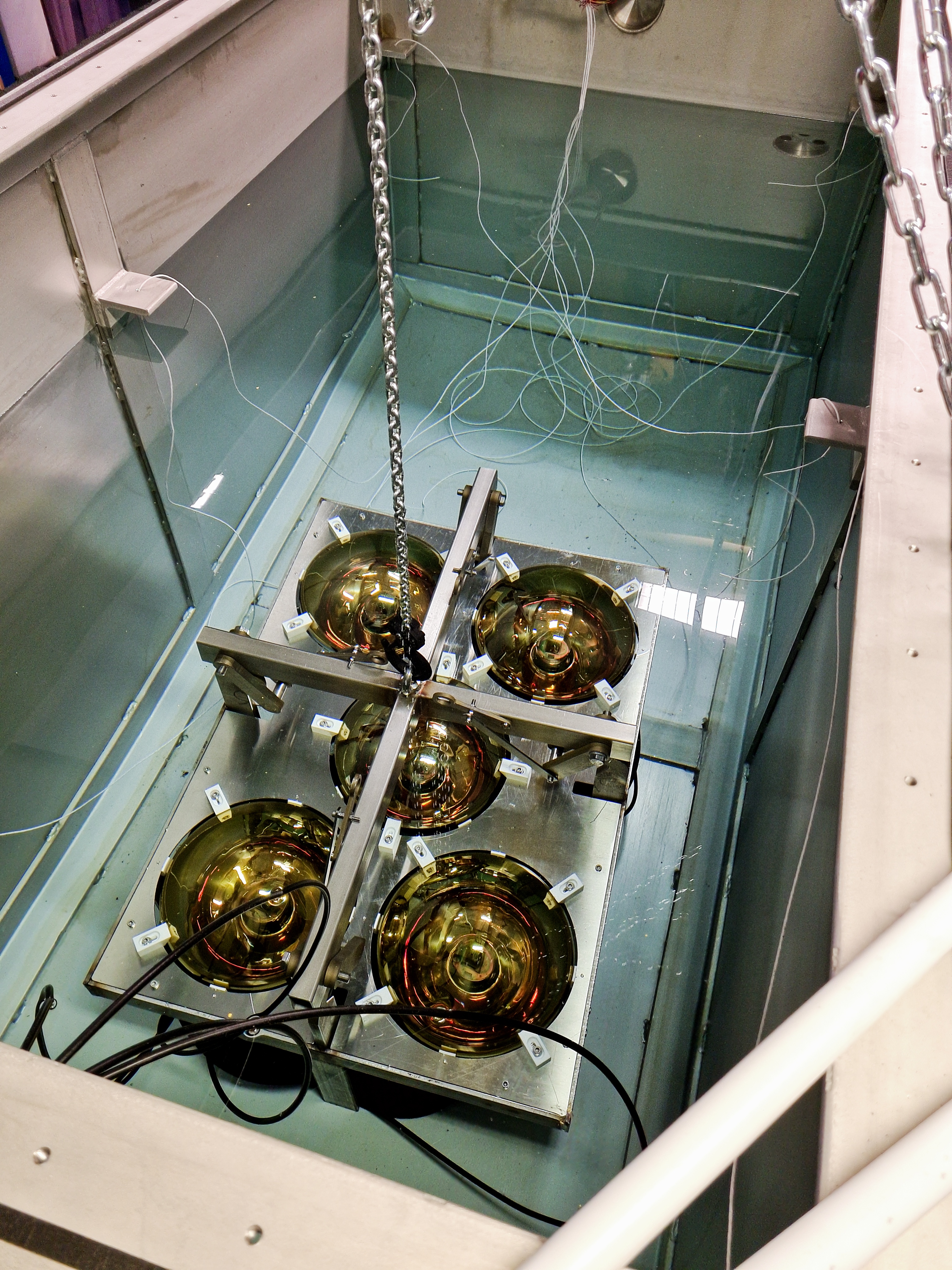}
 \caption{A batch of \acp{pmt} in the holder being lowered into the tank.}
 \label{fig:pmt-holder}
 \end{subfigure}
 \caption{(a) The 2000 L tank used to submerge \acp{pmt} in pure water. The inside of the tank is optically isolated from its surroundings, and the water purity and temperature are controlled. (b) A batch of five \acp{pmt} inside the tank. Covers can be added to each \ac{pmt} individually for further optical isolation.}
 \label{fig:tank-pmts}
\end{figure}

To ensure that no stray light enters the tank, and artificially increases the dark rate, the tank, and lid are covered by layers of black plastic sheets, and the feedthroughs used for \ac{pmt} cables and monitoring of the conditions are also wrapped. \autoref{fig:tank} shows the outside of the tank when it is set up for testing, with \autoref{fig:pmt-holder} showing how the \acp{pmt} are lowered into the tank.

\subsection{Measurement procedure}
\label{subsec:measurements}

To take measurements, a set of up to 5 \acp{pmt} are placed in a holder and lowered into the tank. A lid is placed on the tank and the cables connected to the splitter board after passing through a \ac{shv} to \ac{shv} feedthrough.

To measure the rate of dark pulses, 15 minute runs with the 10 kHz external trigger are used. Each run equates to 1.98 s of live time, or approximately 9 x 10$^6$ waveforms each 220 ns in length. Four runs are taken across a two hour period at each temperature to ensure consistency between runs. The water temperature is adjusted between 7 $^{\circ}$C and 25 $^{\circ}$C in 2.5 $^{\circ}$C steps, except for between 12 $^{\circ}$C and 15 $^{\circ}$C where 0.5 $^{\circ}$C steps are used. This region is chosen to surround the operating temperature of Super-Kamiokande \cite{SK_IV_Calib_2014}.

As measurements of the dark rate are desired, the inside of the tank needs to be kept optically isolated from the tank's surroundings. To confirm no external light is entering the tank, runs are taken with the room lights on and off for each \ac{pmt} deployment. The final data is taken with the room lights off.

\begin{figure}[htb]
\centering
    \includegraphics[width=.6\textwidth]{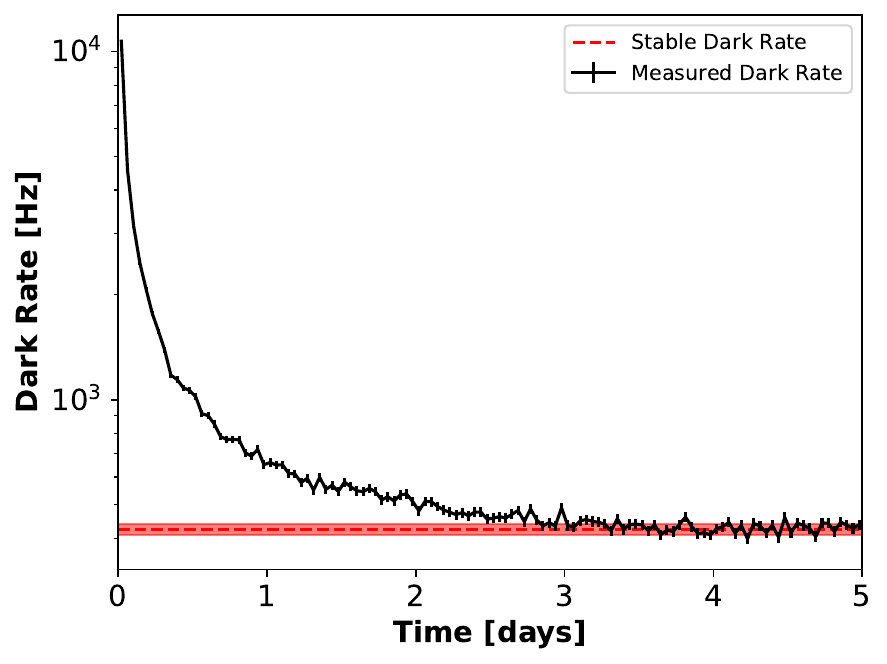}
  \caption[]{The dark rate with time in complete darkness after large light exposure for \ac{pmt}0103 (black). The rate takes around two and a half days to settle at the pre-exposure level of 425 Hz (red, dashed) within error.}
  \label{fig:cooldown}
\end{figure}

When \acp{pmt} are exposed to significant levels of light, they have an increased dark rate for several hours to days after the exposure. This ``cooldown" period is measured by exposing the \acp{pmt} to large levels of light and then recording the dark rate every hour for five days at a fixed temperature. The outcome, in \autoref{fig:cooldown}, shows that after approximately two and a half days the dark rate is at its pre-exposure level. Therefore, the \acp{pmt} are left in total darkness for at least two and a half days, 60 hours, before attempting to take measurements.

Despite light from outside the tank being blocked from entering the system, light can still be produced inside the tank. Cosmic ray muons will produce Cherenkov radiation as they pass through the water. As the tank is on the Earth's surface and near sea level, the rate is approximately 1 cm$^{-2}$ min $^{-1}$ \cite{GADEY2020109209}. This, alongside the large amount of light produced per muon and the amount the light reflects in the steel tank, produces a significant increase in \ac{spe} level pulses observed by the \acp{pmt}. Alongside muons, radioactive backgrounds, in particular from the \ac{pmt} glass, can produce Cherenkov and gamma radiation. It is also possible for light to be produced by the final dynodes in the \ac{pmt} itself \cite{Krall1967}. Most single-photon detecting \acp{pmt} have countermeasures \cite{Wright2017}, but it is possible for photons to escape and trigger that or neighbouring \acp{pmt}.

To combat the light produced inside the system and isolate the thermionic emission and intrinsic dark rate, each \ac{pmt} is individually covered in a black plastic sheet. This allows them to be optically isolated from each other and any light produced in the water. 



\section{Analysis}
\label{sec:analysis}

\subsection{Pulse analysis}
\label{subsec:pulse}

\begin{figure}[!htb]
\centering
    \includegraphics[width=.6\textwidth]{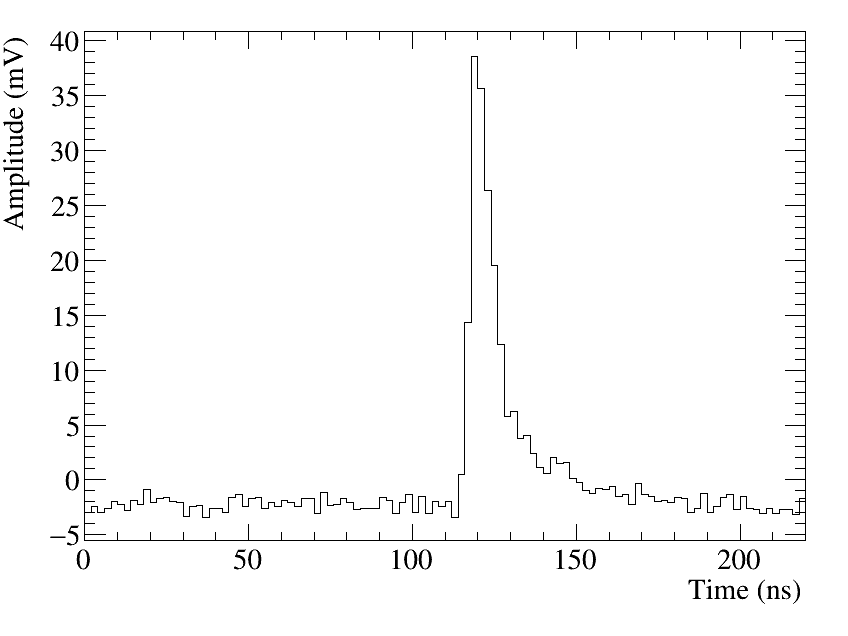}
  \caption[]{Typical \ac{spe} level pulse after 10x amplification. These make up approximately 90\% of the pulses above threshold.}
  \label{fig:pulse}
\end{figure}

An \ac{spe} pulse is characterised as having a mean amplitude of 40 mV after amplification, and a width of around 20 ns where the rise is very sharp (few ns) and the fall is exponential \cite{PMTQA}. An example is shown in \autoref{fig:pulse}. A threshold of 25\% of the mean \ac{spe} amplitude, 10 mV in this case, is used in line with Hamamatsu \cite{Hamamatsu-datasheet}, and a rise time cut of $<$ 20 ns is applied. An upper amplitude limit of 80 mV is applied to remove multiple-photon backgrounds e.g. from cosmological muons. The probability of multiple true dark pulses in a single acquisition window is assumed to be negligible.


\begin{figure}[!htb]
\centering
    \includegraphics[width=.6\textwidth]{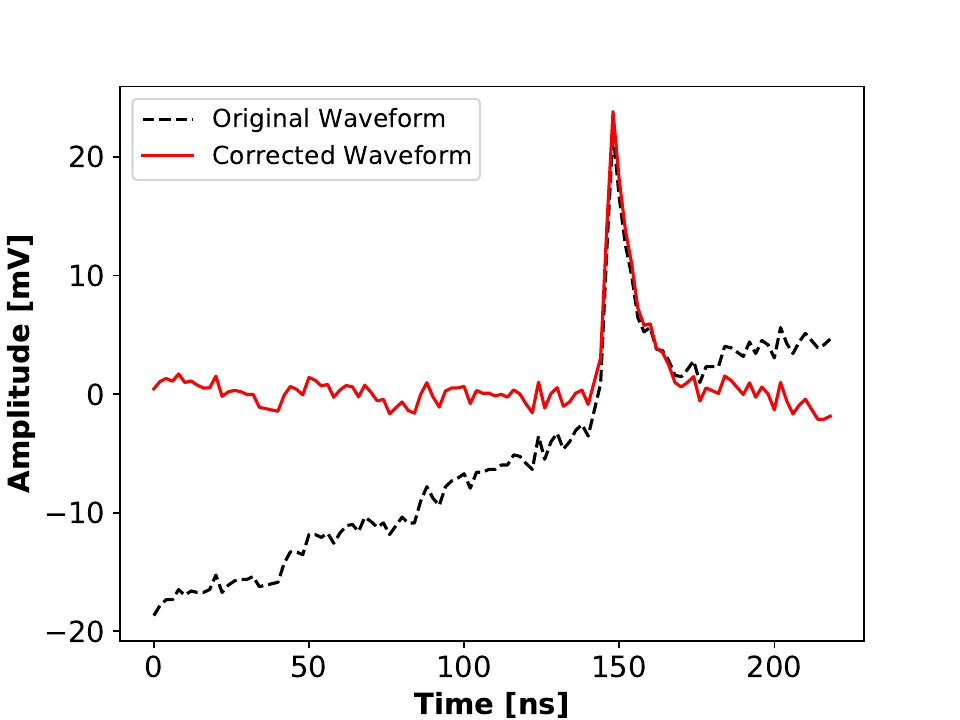}
  \caption[]{The corrections applied to a waveform with a shifted baseline (red, solid), with the original waveform (black, dashed) shown for reference.}
  \label{fig:baseline_correction}
\end{figure}

Despite attempting to match the impedance between the \ac{pmt} electronics and the splitter board, there is still a mismatch due to the circuit board used in the splitter board. This mismatch causes reflections of large pulses in the cables, which causes a large baseline shift within the data acquisition window when these pulses occur just before the external trigger. The system as a whole also contributes to these shifts. The water circulation systems contain several pumps and motors, and the water tank is difficult to shield. As such, the system is susceptible to noise which causes further shifts in the baseline. \Ac{spe} level pulses can be impacted by these baseline shifts, as shown in \autoref{fig:baseline_correction}.


To compensate for this, a baseline correction similar to the Linear Common Mode Suppression algorithm described in \cite{Haefeli:2004ux} is applied. This correction has several steps:
\begin{enumerate}
    \item The mean of the waveform is subtracted.
    \item Any linear slope in the data is corrected for.
    \item Data points more than one standard deviation from the mean are removed.
    \item The mean of the new waveform is subtracted.
    \item Any linear slope in the new waveform is corrected for.
    \item Outlying data points with corrections applied are re-added to the waveform.
\end{enumerate}

This acts to counter the reflections and leave a pulse with a flat baseline centred on 0 mV such as the one in \autoref{fig:baseline_correction}, making the analysis of \ac{spe} level pulses more robust.

\subsection{Thermionic emission}
\label{subsec:thermionic-analysis}

Thermionic emission is modelled using Richardson's law \cite{Richarson1924} in \autoref{eq:richardson},
\begin{equation}
    N = AT^2 \exp\Bigg(\frac{-e\phi}{k_BT}\Bigg).
    \label{eq:richardson}
\end{equation}
Here, $N$ is the rate of electrons emitted from the photocathode by thermionic emission, $A$ is a scaling factor containing a material-specific correction factor, $T$ is the absolute temperature, $\phi$ is the work function, $e$ is the electron charge and $k_B$ is the Boltzmann constant. A constant offset is also applied.

The electrons emitted will produce \ac{spe} level pulses, at the rate $N$. By measuring the dependence of $N$ on $T$, the work function $\phi$ can be obtained for the photocathode and any divergence between the data and Richardson's law for thermionic emission can be observed.
\section{Results}
\label{sec:results}

\begin{figure}[h]
\centering
 \begin{subfigure}[b]{0.475\textwidth}
 \includegraphics[width=\textwidth]{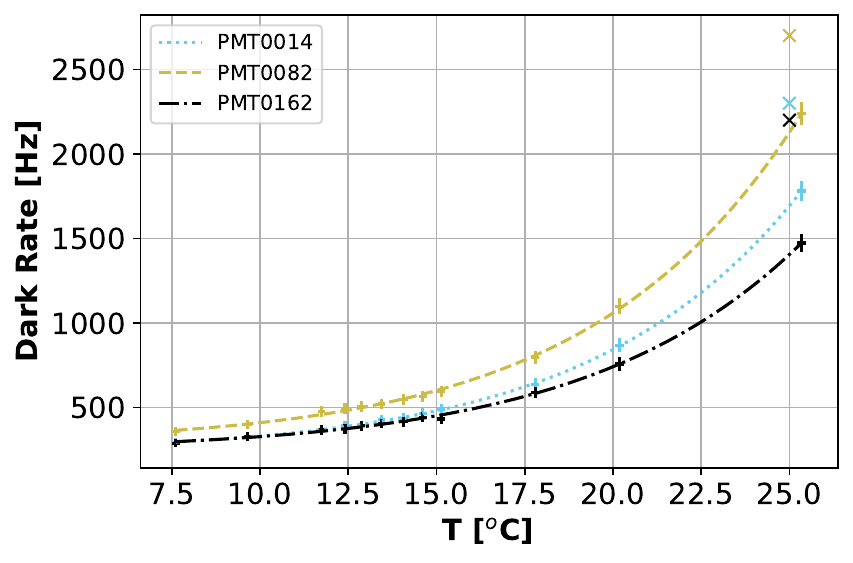}
 \caption{Dark rate vs temperature for covered \acp{pmt}.}
 \label{fig:covered-batch2}
 \end{subfigure}
 \begin{subfigure}[b]{0.475\textwidth}
 \includegraphics[width=\textwidth]{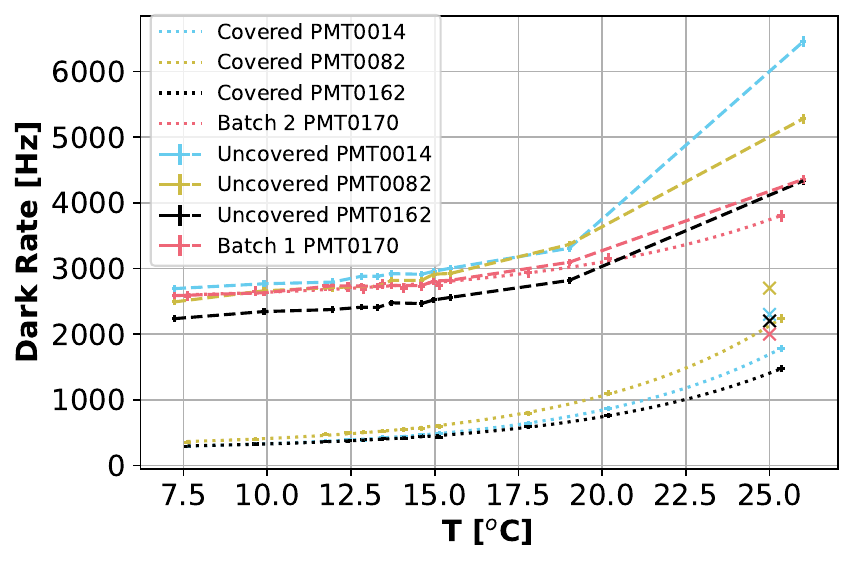}
 \caption{Covered vs uncovered \acp{pmt}.}
 \label{fig:covered_vs_uncovered}
 \end{subfigure}
 \caption{The \ac{spe} level pulse rate for a batch of \acp{pmt}. (a) shows the dark rate with the \acp{pmt} covered. (b) shows a comparison between the \ac{spe} level rates when the \acp{pmt} are optically isolated from their surroundings by a cover and those exposed to the whole system, with PMT0170 kept uncovered to act as a control. Covering the \acp{pmt} lowers the \ac{spe} level count rate significantly, and brings them closer to the measurements made by the vendor \cite{Hamamatsu-datasheet} (``x'' at 25 $^{\circ}$C). The measurements for the covered \acp{pmt} have the fit to Richardson's law superimposed.}
 \label{fig:data-batch}
\end{figure}

The results shown in \autoref{fig:data-batch}, show that Richardson's law fits the data well. The data occasionally has a small plateau-effect below 15 $^\circ$C as the thermionic emission component of dark rate reduces. Results for one batch of three covered \acp{pmt} are shown in \autoref{fig:covered-batch2}. A comparison between the \ac{spe} level pulse rates when covered and uncovered for these \acp{pmt} is shown in \autoref{fig:covered_vs_uncovered}, suggesting that when the \acp{pmt} are uncovered, more than half of the counts come from sources other than dark pulses.



The work functions for the \acp{pmt}, obtained by fitting Richardson's law to the data, match the range measured in air \cite{PMTQA} with values between 1.2 and 1.5 eV.

Runs with the room lights on and off are consistent with each other, confirming no stray light enters the system.

\begin{figure}[htb]
\centering
    \includegraphics[width=.6\textwidth]{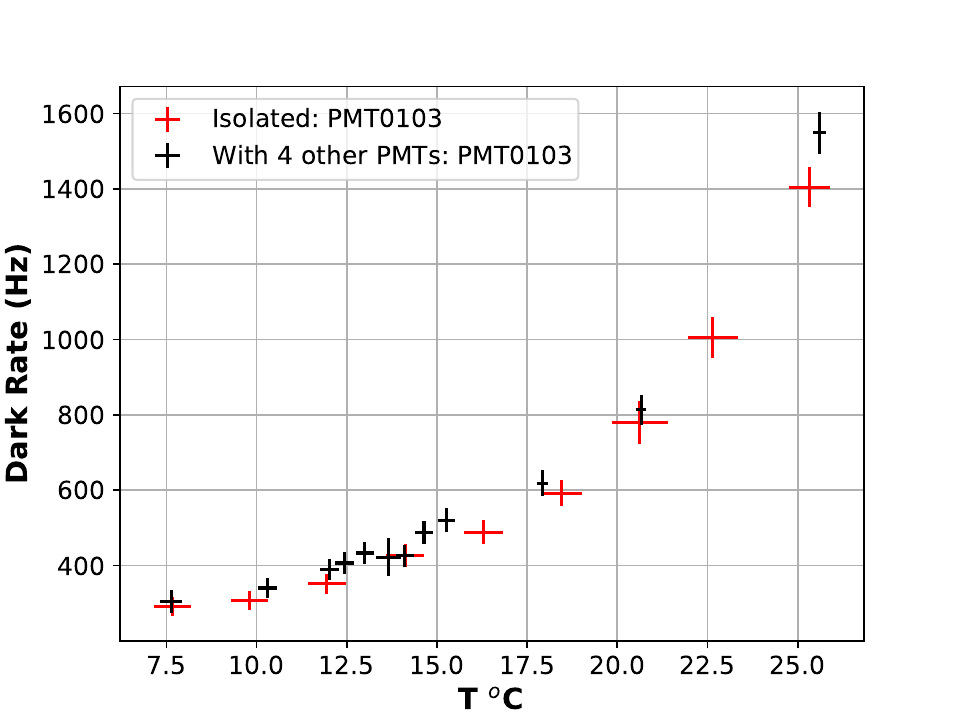}
  \caption[]{The dark rate with temperature for PMT0103 measured when the \ac{pmt} is part of a batch of \acp{pmt} (black) compared to in isolation (red).}
  \label{fig:isolation}
\end{figure}

To confirm the results, a series of tests are conducted on PMT0103 in isolation. To do this, PMT0103 is deployed in the tank on its own. Temperature-dependent measurements of dark rates are taken and compared to the measurements taken from PMT0103 when deployed with other \acp{pmt}. The results, in \autoref{fig:isolation}, show very little difference between the two runs. The \ac{pmt} is also exposed to light before measuring the dark rate every hour for five days which confirms the optical cooldown period of 60 hours is sufficient, with results in \autoref{fig:cooldown}.

\begin{figure}[htb]
\centering
 \begin{subfigure}[b]{0.475\textwidth}
 \includegraphics[width=\textwidth]{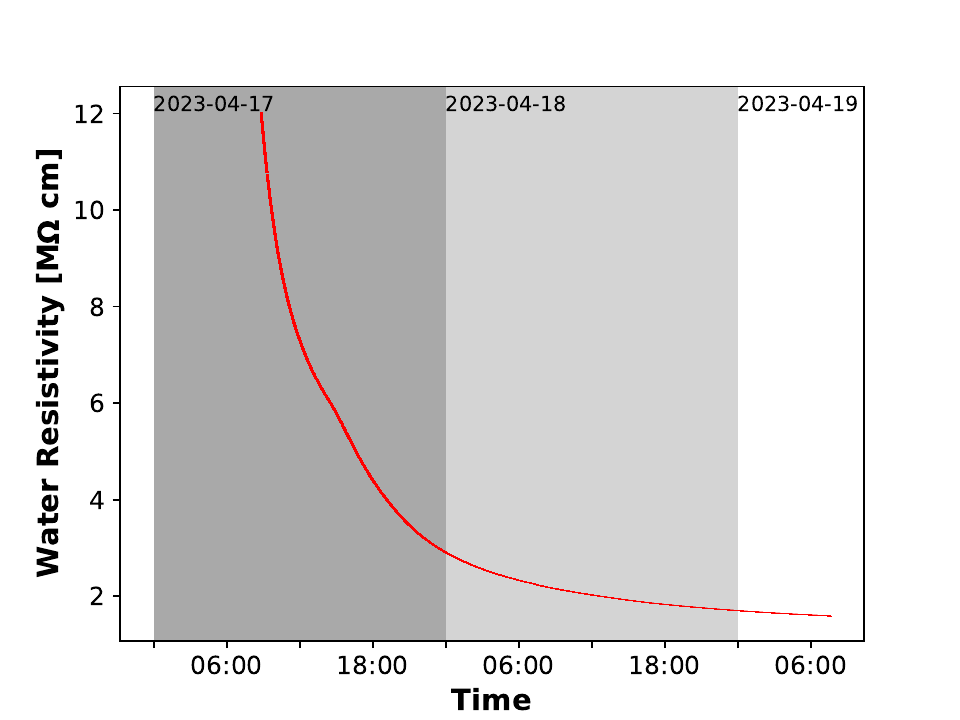}
 \caption{Evolution of water resistivity with time.}
 \label{fig:water-res}
 \end{subfigure}
 \begin{subfigure}[b]{0.475\textwidth}
 \includegraphics[width=\textwidth]{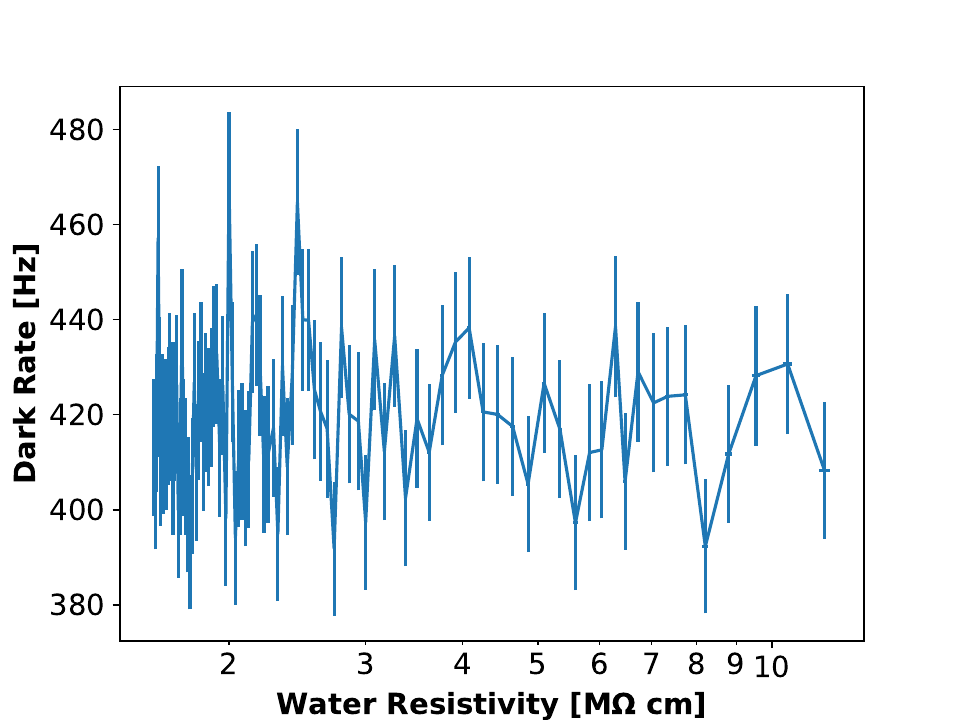}
 \caption{Dark rate with water resistivity.}
 \label{fig:water-res-results}
 \end{subfigure}
 \caption{(a) The water resistivity with time after the deionising cylinder is bypassed. The resistivity drops when the deionising cylinder in the filtration system is bypassed. (b) The dark rate of a covered \ac{pmt} with resistivity. There is no clear connection between intrinsic \ac{pmt} dark rate and water resistivity.}
 \label{fig:water-res-total}
\end{figure}

The impact of water resistivity is tested by bypassing the deionising cylinder and allowing the water resistivity to drop gradually, and taking data every hour during this. The water resistivity and dark rate results, in \autoref{fig:water-res-total}, confirm the resistivity has no clear impact on the intrinsic \ac{pmt} dark rate.
 
\section{Discussion}
\label{sec:dicsussion}

The results of the measurements of temperature-dependent dark rate across thirteen \acp{pmt} are in good agreement with thermionic emission as modelled using Richardson's law in \autoref{eq:richardson} in the 15 $^{\circ}$C to 25 $^{\circ}$C temperature range. However, below 15 $^{\circ}$C there are occasionally small deviations above the expected rate as thermionic emission rates reduce. This suggests that there are non-thermionic emission contributions to the dark rate that become more dominant at lower temperatures. It is seen in \cite{Wright2017,Coates1972} that the temperature-dependent dark rate does not always follow Richardson's law of thermionic emission.

An assay of the \ac{pmt} glass, performed at the \ac{bugs} facility and shown in \autoref{tab:assay}, suggests that the \ac{pmt} glass has a low contribution to the background at approximately 3.4 Bq kg$^{-1}$. The mass of these \acp{pmt} is 1.4 kg, yielding $<$ 5 Bq per \ac{pmt} assuming the glass is the majority of the \ac{pmt} mass. Not all of these events will cause light emission, further lowering the contribution to the overall dark rate.

\begin{table}
    \centering
    \begin{tabular}{ |c|c|cc|c| } 
        \hline
        \bf{Isotope} & \bf{mBq/kg} & \multicolumn{2}{|c|}{\bf{pp}$\mathbf{x}$} \\
        \hline
        $^{238}$U & 660(40) & 54(3) & \multirow{4}{4em}{ppb $^{238}$U}\\
        $^{226}$Ra & 430(20) & 35(1) & \\
        $^{210}$Pb & 610(50) &  50(4) & \\
        $^{235}$U & 30(5) & 50(7) & \\
        \hline
        $^{228}$Ra & 440(20) & 108(5) & \multirow{2}{4em}{ppb $^{232}$Th}\\
        $^{224}$Ra & 370(10) & 89(2) & \\
        \hline
        $^{40}$K & 830(50) & 27(2) & \multirow{1}{4em}{ppm $^{40}$K}\\
        \hline
    \end{tabular}
    \caption{\ac{pmt} glass assay results from data taken at the \ac{bugs} facility using a Mirion (Canberra) BE6530 BEGe-type HPGe detector. Uncertainties in mBq/kg are 1 $\sigma$ statistical errors only. True coincident summing effects are not accounted for, and introduce an additional error of up to $\mathcal{O}$(10\%). For comparison only, the $^{235}$U value is given as a $^{238}$U ppb equivalent where conversion is performed using the standard $^{nat}$U activity ratio for $^{238}$U: $^{235}$U of 21.6 \cite{IAEA2023}.}
    \label{tab:assay}
\end{table}

Measurements of the \ac{spe} level pulse rate with changing water resistivity confirm that the water resistivity has negligible impact on the intrinsic \ac{spe} level pulse rate.

The largest non-thermal contribution to \ac{spe} level counts in a liquid medium is light. Despite the system being optically isolated from its environment, the liquid fill allows light production via particle interactions from cosmic ray particles and radioactive decays. Cosmic ray muons are incident on the tank at an expected rate of the order several hundred Hz as the tank has a 2 m$^2$ top surface area, and a further 6 m$^2$ on the sides. This light is also likely to be reflected, which will further increase the number of photons hitting \acp{pmt}, due to the steel tank. Without simulations, it is hard to quantify the expected number of hits from muon-produced photons, but it expected to be a significant contributor to the total number of counts observed by the \acp{pmt}. Along with the direct detection of photons, the consistent light exposure leads to a prolonged increase in \ac{pmt} activity leading to a further increase in the \ac{spe} level pulse rate. As such, simply vetoing cosmic ray muons and accounting for radioactive backgrounds might not allow all contributions to the \ac{spe} pulses due to light to be removed. Beyond light production in the tank, high gain \acp{pmt} are known to produce light in the final dynodes around the anode \cite{Krall1967}. This light can trigger both the \ac{pmt} emitting the light and its neighbours. By optically isolating \acp{pmt} from the system and their neighbours, the \ac{spe} level pulse rate drops by a factor of $\sim$ 2.5 as shown in \autoref{fig:covered_vs_uncovered}.

A comparison between dark rate measurements made in water and air (both in \cite{PMTQA} and by the vendor \cite{Hamamatsu-datasheet}) at 25 $^\circ$C is presented in \autoref{tab:results-25deg}. The measurements in air show good agreement. The water measurements agree with expectation when the time in darkness before data taking is accounted for. The measurements made in air are after approximately 18 hours in complete darkness, whereas those in water are after at least 60 hours. The dark rate reduces significantly over this time. For PMT0103, the dark rate after 18 hours is 1.8 times the rate after 60 hours, as shown in \autoref{fig:cooldown}.

\begin{table}[]
    \centering
    \begin{tabular}{|c|c|c|c|}
        \hline
        \multirow{2}{2.5em}{\bf{PMT ID}} & \multicolumn{3}{|c|}{\bf{Dark Rate at 25 $^\circ$C [Hz] (after time in dark)}} \\
         & \bf{Water (60 hours)} & \bf{Air (18 hours)} & \bf{Vendor} \\
        \hline
        14 & 1,708 & 2,409 & 2,300\\
        82 & 2,149 & 2,793 & 2,700\\
        162 & 1,396 & 2,141 & 2,200\\
        3 & 1,804 & 4,401 & 3,600\\
        112 & 1,730 & 1,855 & 1,500\\
        159 & 1,235 & 2,489 & 2,000\\
        26 & 1,329 & 3,702 & 3,600\\
        104 & 3,276 & 1,667 & 1,500\\
        143 & 865 & 1,439 & 1,400\\
        155 & 3,535 & 4,193 & 5,000\\
        15 & 1,397 & 2,259 & 2,200\\
        29 & 1,487 & 1,729 & 1,900\\
        103 & 1,428 & 1,759 & 1,600\\
        148 & 954 & 1,892 & 2,100\\
        \hline
    \end{tabular}
    \caption{The dark rate measured at 25 $^\circ$C in water, air \cite{PMTQA} and by the vendor (air) \cite{Hamamatsu-datasheet}. The tests in water were conducted after at least 60 hours in complete darkness, the other tests after approximately 18 hours. Uncertainties are 1 - 4\% for the water and air measurements.}
    \label{tab:results-25deg}
\end{table}

This is significant for detector application, as it suggests that there is no intrinsic difference in \ac{pmt} dark rate when the medium of deployment changes. Therefore, any differences seen in \ac{pmt} rates when in a liquid are likely to be due to external contributions such as light production in the system. In the case of LUX-ZEPLIN \cite{Wright2023}, the raised pulse rates that occur after the tank is opened and the water resistivity drops is likely related to contaminants in the water or optical cooldown after light exposure, and not an intrinsic effect of water resistivity on the \acp{pmt}.
\section{Conclusion}
\label{sec:conclusion}

The temperature-dependent dark rate of Hamamatsu R7081-100 10" \acp{pmt} has been characterised when submerged in water, and shows a good match to the measurements taken in air \cite{PMTQA} and by the vendor \cite{Hamamatsu-datasheet}. Richardson's law of thermionic emission fits the data well, especially in the 15 $^\circ$C to 25 $^\circ$C range, with other contributions becoming more noticeable at lower temperatures.

The impact of light on the \ac{spe} level pulse rate is significant, and simply using a light tight water tank is not sufficient to remove all light. This is due to light production in water from cosmic ray muons, radioactive backgrounds and possibly the \acp{pmt} themselves. The impact of radioactivity from individual \acp{pmt} and water resistivity are shown to have a negligible effect on the intrinsic dark rate, but exposure to light has a lasting impact. The source of raised \ac{spe} level pulse rates in liquid detectors is likely to be external to the \acp{pmt}.

It can be concluded that intrinsic \ac{pmt} dark rate is consistent across different media, and that characterising a \ac{pmt} in air is sufficient, based on the comparison of temperature-dependent dark rates measured in air and water when all light is removed.

\section*{Acknowledgments}
The authors would like to express our gratitude to Matt Thiesse at the University of Sheffield for his design and construction of the water tank used for this testing. We would also like to express our thanks for his regular review of the concepts and methods involved in this work.

The authors would also like to thank Paul Scovell and the team at the \ac{bugs} facility at STFC's Boulby Underground Laboratory for their assay of the \ac{pmt} glass.

Finally, we would like to acknowledge the work done by Tom Shaw and Steve Quillin of the Atomic Weapons Establishment in the United Kingdom. Tom designed the first version of the electronics setup used, whilst Steve was involved in the early stages of the waveform analysis.

\bibliographystyle{JHEP}
\bibliography{biblio.bib}
\end{document}